\documentclass[aps,pre,twocolumn,amsfonts,amssymb,amsmath,showpacs]{revtex4}
\usepackage[dvips]{graphicx}

\newcommand{\eg}[1]{{\it e.g.\/}\ifx#1.\else\expandafter#1\fi}
\newcommand{\ie}[1]{{\it i.e.\/}\ifx#1.\else\expandafter#1\fi}

\renewcommand{\@}{\partial}
\newcommand{\const}{\mathrm{const}}
\newcommand{\Df}[2]{\displaystyle{\frac{\d #1}{\d #2}}}
\renewcommand{\d}{\mathrm{d}}
\newcommand{\df}[2]{\frac{\partial #1}{\partial #2}}
\newcommand{\ddf}[2]{\frac{\partial^2 #1}{\partial #2^2}}
\newcommand{\eq}[1]{\eqref{#1}}
\newcommand{\Mx}[1]{\left[\begin{array}{cc}#1\end{array}\right]}
\newcommand{\mx}[1]{\mathbf{#1}}
\newcommand{\Real}{\mathbb{R}}
\renewcommand{\Re}{\mathrm{Re}}

\newcommand{\D}{\mx{D}}                 
\newcommand{\Dr}{\D_r}                  
\newcommand{\DP}{D_{\P}}                
\newcommand{\DZ}{D_{\Z}}                
\newcommand{\dt}{\Delta t}              
\newcommand{\dx}{\Delta x}              
\newcommand{\F}{\mx{F}}                 
\newcommand{\Fr}{\F_r}                  
\newcommand{\f}{\mx{f}}                 
\newcommand{\hm}{h_{-}}                 
\newcommand{\hp}{h_{+}}                 
\newcommand{\I}{\mx{I}}                 
\renewcommand{\k}{k}                    
\newcommand{\lyap}{\lambda}             
\renewcommand{\u}{\mx{u}}               
\newcommand{\uo}{\mx{u}_o}              
\newcommand{\ur}{\mx{u}_r}              
\renewcommand{\v}{\mx{v}}               
\renewcommand{\P}{u}                    
\renewcommand{\Pr}{\P_r}                
\newcommand{\Pc}{\P_*}                  
\newcommand{\Z}{v}                      
\newcommand{\Zr}{\Z_r}                  

\newcommand{\Fig}[1]{Fig.~\ref{fig:#1}}
\newcommand{\Figs}[1]{Figs.~\ref{fig:#1}}
\newcommand{\fig}[1]{fig.~\ref{fig:#1}}
\newcommand{\figref}[1]{\ref{fig:#1}}
\newcommand{\dblfigure}[3]{
  \begin{figure*}[tb!]
  \includegraphics{#1}
  \caption[]{#2}
  \label{fig:#3}
  \end{figure*}
}
\newcommand{\sglfigure}[3]{
  \begin{figure}[tb!]
  \includegraphics{#1}
  \caption[]{#2}
  \label{fig:#3}
  \end{figure}
}

\begin{document}
\title{Spontaneous traveling waves in oscillatory systems with cross diffusion}
\author{V. N. Biktashev}
\affiliation{Department of Mathematical Sciences, 
  University of Liverpool, Liverpool L69 7ZL, UK }
\author{M. A. Tsyganov}
\affiliation{
  Institute of Theoretical and Experimental Biophysics, 
  Pushchino, Moscow Region, 142290, Russia}
\date{\today}
\begin{abstract}
  We identify a new type of pattern formation in spatially distributed
  active systems.  We simulate one-dimensional two-component systems
  with predator-prey local interaction and pursuit-evasion taxis
  between the components. In a sufficiently large domain, spatially
  uniform oscillations in such systems are unstable with respect to
  small perturbations. This instability, through a transient regime
  appearing as spontanous focal sources, leads to establishment of
  periodic traveling waves. The traveling waves regime is
  established even if boundary conditions do not favor such solutions.
  The stable wavelength are within a range bounded both from above and
  from below, and this range does not coincide with instability bands
  of the spatially uniform oscillations.
\end{abstract}
\pacs{%
  87.10.+e
, 02.90.+p
}
\maketitle

\section{Introduction}

Dissipative structures, \ie\ patterns in spatially extended
  systems away from equilibrium have been intensively studied for many
  decades. A very comprehensive review can be found in
  \textcite{Cross-Hohenberg-1993}; results obtained since then would
  probably require an even more extensive review.  A very popular
  class of mathematical models is the reaction-diffusion systems with
  diagonal diffusion matrices. There have been numerous indications
  that non-diagonal elements in diffusion matrices, \ie\
  cross-diffusion, can lead to new nontrivial effects not observed in
  classical reaction-diffusion systems, \eg\ \emph{quasi-solitons} in
  systems with excitable reaction part
  \cite{QS1,QS2,QS3,QS5,QS6}. However oscillatory systems are more prevalent
  than excitable and nontrivial effects of cross-diffusion in oscillatory systems have not been 
  studied yet. Here we consider an example where the
  reaction part of the system is dissipative while the diffusion part
  is not. We describe spontaneously generated periodic waves, and
  identify the features of these waves that indicate that we are
  dealing here with a phenomenon not seen before.

A general formulation of a reaction-diffusion system with nonlinear diffusion is
\begin{equation}
  \df{\u}{t} = \f(\u) + \nabla(\D(\u)\nabla\u), \qquad 
  \u,\f\in\Real^N, \; \D\in\Real^{N\times N}. \label{RD}
\end{equation}
Both the reaction term $\f(\u)$ and the diffusion term $\nabla(\D(\u)\nabla\u)$
in the right-hand side represent dissipative processes, which for
diffusion implies that matrix $\D\in\Real^{N\times N}$
is positive (semi-)definite, typically diagonal
with non-negative elements. A huge amount of results have been
obtained about pattern formation described by such models. However,
many physical situations lead to non-diagonal elements in $\D$, \ie\
cross-diffusion (see \eg\ discussions in \cite{UFN07,Vanag-Epstein-2009}).
Some such situations may be adequately
described by $\D$ whose eigenvalues have zero real part, \eg\ when the
self-diffusion of components is negligible. In such cases reaction
part is dissipative and the ``diffusion'' part is not. Physical
consequences of such ambivalence are little understood yet.

Cross diffusion has been seen to produce interesting phenomena, such
as fronts, pulses and stationary periodic structures (see \eg\
\cite{del-Castillo-Negrete-etal-2002,Chung-Peacock-2007}
among many other works), however phenomenologically similar regimes
are known in reaction-diffusion systems, too.

In a recent series of works we have described unusual phenomena, such
as quasi-solitons and their variations, in excitable systems in which
linear or nonlinear cross-diffusion was added to or replaced
self-diffusion (see \eg\ \cite{QS1,QS2,QS3,QS5,QS6}). The ability of a medium
to conduct solitary waves is stipulated by its excitable kinetics
described by the reaction term $\f(\u)$, whereas specifics of their
interaction are also due to the cross-diffusion terms. However,
excitability is a relatively exotic, albeit very important, type of
behaviour compared to oscillations. For instance, in population
dynamics, plausible excitable predator-prey models have been proposed
\cite{Truscott-Brindley-1994} but we are not aware of reliable observations
of natural systems described by such models. On the other hand,
oscillatory behaviour in predator-prey systems is textbook material
\cite{Murray-2002,Britton-2003} and there are plentiful observational
data on traveling waves in cyclic populations
\cite{Sherratt-Smith-2008}.

Solitary waves in oscillatory systems are not feasible, and it is not
clear what new features cross-diffusion may impose.

\dblfigure{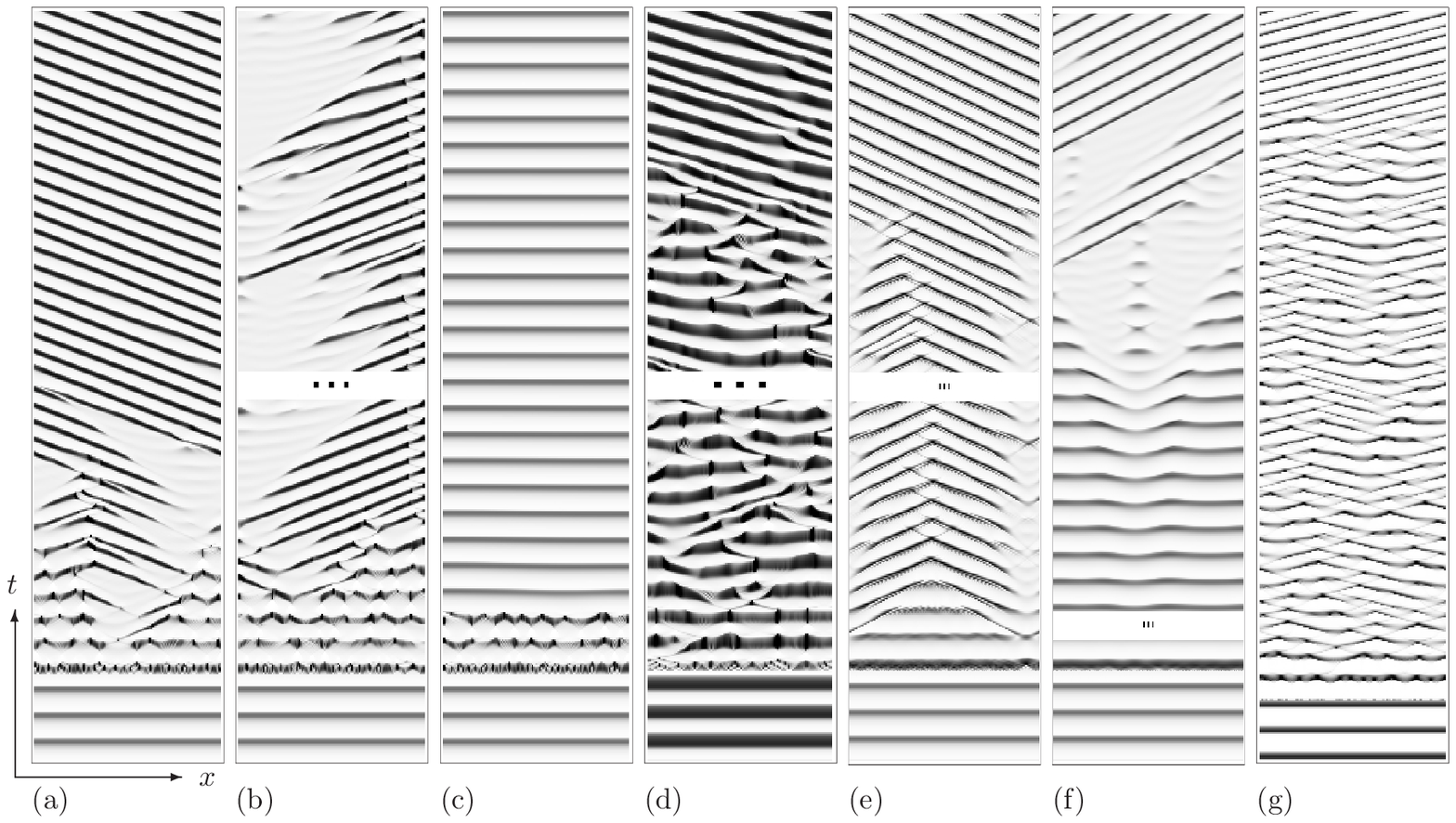}{
  Different regimes resulting from random perturbaton of uniform
  oscillations. 
  Shown are density plots: space $x$ is horizontal, time $t$ is
  vertical increasing upwards, $\P=1$ corresponds to black and $\P=0$
  corresponds to white.
  (a) TB model, $L=15$, taxis ($\hm=1$, $\hp=0$, $\DP=\DZ=0$), periodic boundary conditions, $t\in[0,2500]$.
  (b) Same, no-flux boundary conditions, $t\in[0,1200]\cup[43800,45000]$.
  (c) Same as (a) except $\DP=\DZ=0.05$, $\hp=\hm=0$. 
  (d) Same as (a) except $w=0.07$, $L=10$, $t\in[0,1200]\cup[36300,37500]$.
  (e) Same as (a) except $\hp=0.1$, $L=50$, $t\in[0,1200]\cup[5000,6200]$.
  (f) Same as (a) except $\hp=0.1$, $\DP=\DZ=0.02$, $L=50$, $t\in[0,400]\cup[1900,3900]$.
  (g) RM model, $\hm=1$, $\hp=\DP=\DZ=0$, $L=25$, $t\in[0,2500]$.
}{density}

The purpose of this communication is to describe new phenomena we have
observed in oscillatory systems with ``pursuit-evasion'' nonlinear
cross-diffusion interaction between the components. 

\section{The models}


We consider two predator-prey models with cross-diffusion terms of
``pursuit-evasion'' mutual taxis,
\begin{eqnarray}
\df{\P}{t} &=& f(\P,\Z) + \DP \ddf{\P}{x} + \hm \df{}{x}\left(\P\df{\Z}{x}\right) , \nonumber\\
\df{\Z}{t} &=& g(\P,\Z) + \DZ \ddf{\Z}{x} - \hp \df{}{x}\left(\Z\df{\P}{x}\right) , \label{RT}
\end{eqnarray}
for $(x,t)\in[0,L]\times[0,t_{\max}]$ 
for two reaction kinetics, the Truscott-Brindley (TB) model
\cite{Truscott-Brindley-1994} 
\begin{eqnarray}
f(\P,\Z) &=& \beta \P(1-\P) - \Z\P^2/(\P^2+\nu^2), \nonumber \\ 
g(\P,\Z) &=& \gamma \Z\P^2/(\P^2+\nu^2) - w\Z ,      \label{TB} 
\end{eqnarray}
where $\beta=0.43$, $\nu=0.053$, $\gamma=0.1$ and $w=0.055$ unless
stated otherwise, and the
Rosenzweig-MacArthur (MA) model
\cite{Rosenzweig-MacArthur-1963,Britton-2003,Sherratt-Smith-2008}
%
\begin{eqnarray}
f(\P,\Z) &=& \beta \P(1-\P) - \Z\P/(\P+\nu), \nonumber \\ 
g(\P,\Z) &=& \gamma \Z\P/(\P+\nu) - w\Z , \label{RM} 
\end{eqnarray}
where $\beta=1$, $\nu=0.3$, $\gamma=0.15$ and $w=0.03$ unless stated
otherwise. 
Here $\P$ represents
prey, $\Z$ predators, the term with $\hp$ describes pursuit of prey by
predators and the term with $\hm$ describes evasion of predators by prey.
The simulation were done on an interval $x\in[0,L]$ with 
periodic or Neumann boundary conditions for both
components, using forward Euler stepping in time,
center differences for the diffusion terms and upwind difference for
the taxis terms, see \textcite{QS2} for details and justification.
Except where stated otherwise, we used discretization
steps $\dx=0.1$ and $\dt=4\cdot10^{-4}$.

\section{Numerical observations}

\sglfigure{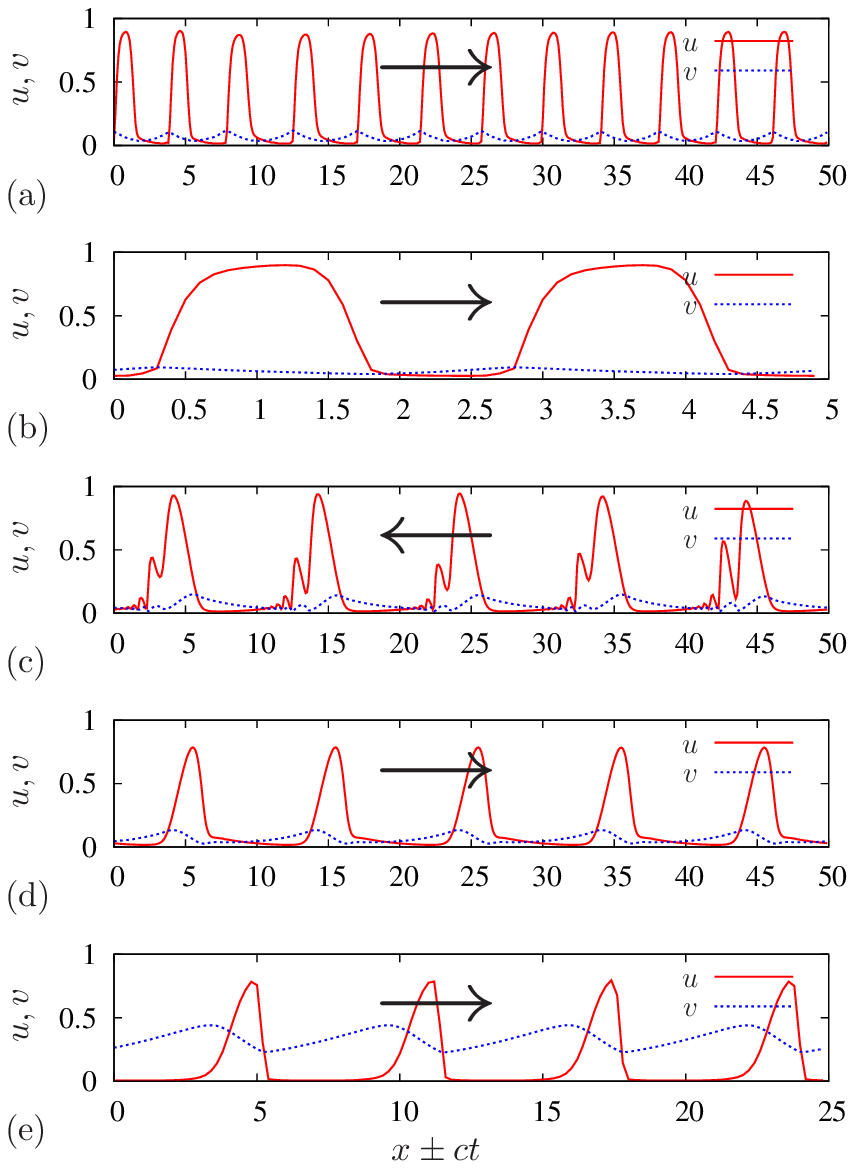}{ (color online)
  Examples of profiles of spontaneously established periodic
    waves. 
  Shown are dependencies of $\P$ and $\Z$ on $x$ at fixed $t$,
  and direction of propagation by arrows. 
  Parameters are the same as in \fig{density} except 
  interval lentgh $L$, specifically, 
  (a) as in \fig{density}(a),
  (b) as in \fig{density}(d), 
  (c) as in \fig{density}(e), 
  (d) as in \fig{density}(f) and
  (e) as in \fig{density}(g).
}{profs}

\Figs{density} and \ref{fig:profs} illustrate the phenomenon of the
spontaneous onset of periodic waves.  Starting from arbitrary spatially-uniform
intial conditions at $t=0$, after a transient allowed to establish
uniform oscillations, perturbations were introduced and subsequent
evolution observed. The perturbation was introduced at half of the
grid points chosen randomly, where at $t=300$ the values of $\P$ were
replaced by randomly chosen numbers in the interval between $0.15$ and
$0.45$. \Fig{density} shows space-time density plots and \fig{profs} illustrate
selected profiles of the emerging wavetrains. 

In the TB model with periodic boundary conditions (\fig{density}(a)),
after a ``random'' transient lasting two or three bulk oscillation
periods, patterns start to emerge: waves start ``from nowhere'' and
annihilate upon collision with other such waves. After a few periods
of such collisions, the waves propagating leftwards win over and a
periodic wavetrain establishes which then persists.  Different seeds
in the random number generator produce solutions differing in details
but always leading to periodic trains, leftward and rightward
propagating with equal probability (compare 
density plots and wave profiles in \fig{density} and \fig{profs},
which corresponded to different simulations with the same parameter sets). 

Impenetrable boundaries do not allow periodic wavetrain solutions;
however the tendency to establish periodic wavetrains is observed even
then. In \fig{density}(b) rightward propagating waves win over. Their impact
with the right boundary $x=L$ is with partial reflection, when the
reflected wave is weak and soon decays; note that this behaviour is
typical for collision of solitary excitation waves in such systems
\cite{QS2}. The left boundary has a quenching effect, but at a
distance from it waves emerge spontaneously. This distance varies
irregularly, indicating that spontaneous generation of waves is
associated with an instability, thus sensitive dependence on initial
conditions and probably chaotic dynamics. This irregular pattern
persists for a long time.

This behaviour is in a contrast with a system with the same
kinetics but pure diffusional spatial terms: in \fig{density}(c), similar
initial random perturbations lead very quickly to re-establishment of
spatially uniform oscillations. 

The parameters used in \fig{density}(a) are close to the boundary of
the oscillatory regime in the TB model (achieved \eg\ at
$w\approx0.053$ with other parameters fixed). When parameters are
further into the oscillatory region, spontaneous generation of
periodic wavetrains is still observed, although the transient period
of spontaneous wavelet generations and collisions lasts longer, see
\fig{density}(d). 

We have also found that prevalence of the ``evasion'' taxis ($\hm$ coefficient) helps
generation of periodic trains, but $\hp=0$ is not necessary, and such
generation can be observed with the ``pursuit'' taxis present as well,
see \fig{density}(e). 

Spontaneous generation of periodic trains is observed in the RM model
as well, see \fig{density}(f). 

\sglfigure{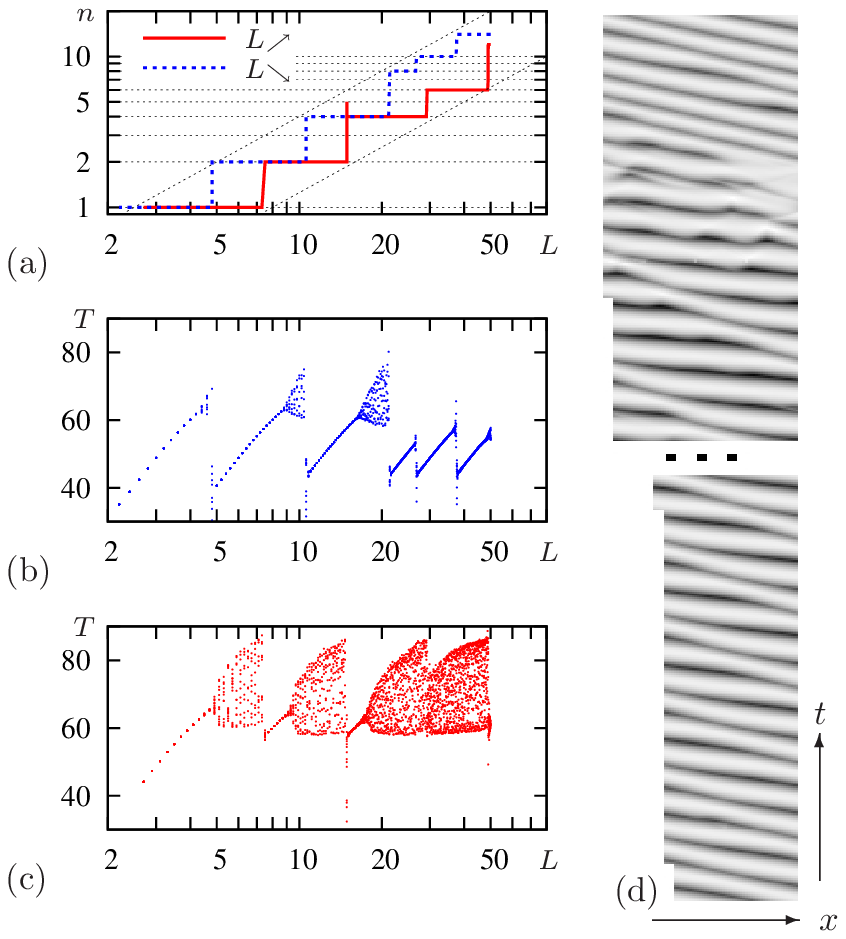}{ (color online)
  Variability of the wavetrains at changing $L$. 
  (a) Number of waves in the interval $[0,L]$ as $L$
  gradually increases (by $0.2$ every 2000 time units, solid red
  line) and decreases (by $0.2$ every 1000 time units, dashed blue
  line). Oblique dashed lines: $n=L/2.5$ and $n=L/8$ to guide the eye.
  (b) Wave periods measured at a point, as function of
  $L$ as it decreases. 
  (c) Same, as $L$ increases. 
  (d) Density plots of two episodes of simulation 
  of 1200 time units duration each,
  with $L$ increasing
  by $0.2$ every 1000 time units. 
  Lower episode: soft transition from steady 1-wave
  solution to modulated 1-wave solution ($L: 4.8 \to 5.6$). 
  Upper episode: subsequent
  sudden transition from modulated 1-wave solution to
  a steady 2-wave solution ($L:7.2 \to 7.6$). 
}{lchange}

The spontaneously emerging periodic wavetrains typically had
wavelengths in a limited range. To check whether this is dictated by
initial conditions or is due to limitations of the system, we
performed simulations in a circle, \ie\ an interval with periodic
boundary conditions, of a slowly changing length $L$. We started from
an established propagating wave in a circle.  Then we changed the
length $L$ of the circle in small steps, allowing sufficient time
between the steps for the waves to adjust. During the simulation we
monitored the number of waves $n$, determined via the number of points
where $\P$ crossed the level $\P=\Pc=0.2$, and the periods $T$ defined as
intervals between $\P$ crossing the level $\P=\Pc$ in the positive
direction. Results of one such simulation are shown in \fig{lchange}.

The number of waves $n$ in the interval did not remain constant
(\fig{lchange}(a)), but spontaneously adjusted so as to keep the
average wavelength within certain limits: between approximately 2.5 and 8 in the
simulation shown. This number was not a unique function of the
interval length: changing $L$ upwards and downwards produced different
dependencies $n(L)$, \ie\ we have hysteresis.  Simulations at slower
rate of change of $L$ slightly changed the $n(L)$ dependencies but the
hysteresis stayed.  Near the transition points where $n$ changed the
value, the propagation of the waves was non-stationary, and was always
for $L$ just below the transitional value, whether it was decreasing
(\fig{lchange}(b)) or increasing (\fig{lchange}(c)). Increasing $L$
had a noticeably more destabilizing effect than decreasing.

The nature of the non-stationary solutions is illustrated by the
density plots shown in \fig{lchange}(d). Starting from an $n=1$
solution, an increase of $L$ above the value of $L\approx5$ leads to
an instability of the steady propagating wave solution. This is a
soft, Eckhaus-type
instability and leads to a mild modulation of the wave, producing
a seemingly two-periodic motion. The amplitude of the modulations
grows as $L$ increases, until at $L=7.6$ a qualitative transformation
occurs. A gap between the wave and its own copy around the circle
grows so big that at a certain moment it is sufficient to allow
spontaneous generation of another wave, leading to an $n=2$
solution. This solution is steady, \ie\ propagates without
modulations, until $L$ grows so big it in turn becomes unstable etc.

\section{Preliminary theoretical considerations}

Substantial theoretical analysis of the phenomenon of the spontaneous
traveling periodic waves is beyond the scope of this communication.
Here we consider one naive approach and then some known pattern
formation mechanisms, which a priori might look relevant to this
phenomenon, only to eliminate them, as not providing a satisfactory
explanation. We will refer to the
historical review by 
\textcite[p.~870]{Cross-Hohenberg-1993} (CH for brevity), and to
a recent symmetry based classification of instabilities and
bifurcations of periodic dissipative waves and structures given by
\textcite[p.~2680]{Rademacher-Scheel-2007} (RS for
brevity).

\paragraph{It is not captured by lambda-omega approach}

The simple class of two-component reaction-diffusion systems
introduced by \textcite{Kopell-Howard-1973} and called
``lambda-omega systems'', and closely related to the complex Ginzburg-Landau
equation, allows exact solutions in the form of periodic waves. It has
offered qualitative insight in many nonlinear wave phenomena,
including periodic waves in cyclic
populations~\cite{Sherratt-Smith-2008}. However, it does not seem to be
helpful in our present case. The modification of the lambda-omega
system, corresponding to the choice of signs of taxis terms in \eq{RT}
is
\begin{equation}
  \df{z}{t} = (\Lambda(|z|) + i\Omega(|z|)) z - i\nabla^2z
\end{equation}
where $z$ is a complex dynamic variable representing $\P+i\Z$, 
and the purely imaginary diffusivity here corresponds to the absence of
self-diffusion, $\DP=\DZ=0$. Then the periodic traveling wave ansatz
$z=a\,\exp[i(\omega t-\k x)]$, $a,\omega,\k\in\Real$, gives the finite system
\[ \Lambda(a) =0 , \qquad \omega = \Omega(a) - \k^2, \]
\ie\ all waves have the same amplitude which is a root of $\Lambda()$,
and exist for all wavelengths $\k$ rather than in a finite interval.
Stability analysis and consideration of nonzero self-diffusion do
not help either. 

\paragraph{It does not emerge via  Turing mechanism.} The instability of
spatially-uniform solutions in favour of non-oscillatory,
spatially-periodic solutions with periods in a finite range is, of
course, a defining feature of the Turing patterns, called just so by
RS and classified as type $I_s$ in CH nomenclature. Cross-diffusion
can provide an alternative to the original Turing's short range
inhibition - long range activation condition.  Indeed Turing-type
instabilities and spontaneously occurring, self-supporting
time-stationary spatially periodic patterns have been observed in
locally multistable systems with cross-diffusion
\cite{del-Castillo-Negrete-etal-2002}.  Our present observations are
different in that here we are dealing with time-oscillating phenomena
not just space-oscillating.

\paragraph{It does not emerge via Turing-Hopf mechanism.}

A Hopf bifurcation of the spatially uniform equilibrium at a
nonzero wavelength, is called ``Hopf'', ``oscillatory Turing'' and
``Turing Hopf'' instability by RS, classifed as type $I_o$ in CH
nomenclature, and also known as short-wave instability or
finite-wavelength instability. It can lead to stable periodic
propagating waves, in lasers, fluid convection and reaction-diffusion
models
\cite{Swift-Hohenberg-1977,Haken-1983,Livshits-1983,Lega-etal-1994}.
In reaction-diffusion context, such waves have been observed
experimentally and in simulations in populations and BZ reaction
\cite{Mendelson-Lega-1998,Vanag-Epstein-2002}.  However, the standard
way such instability occurs in systems \eq{RD} implies existence of an
equilibrium that is stable with respect to spatially uniform
perturbations, which we do not have here, and it only can occur if
$N\ge3$ whereas we have only two components, $\P$ and $\Z$.

Specifically, for $\u(x,t)=\ur+\v\,e^{\lambda t+i\k x}$ where 
$\ur=(\Pr,\Zr)$ is the spatially uniform equilibrium and $|\v|\ll1$, 
we have the characteristic equation
\[
  \det\left( \Fr - \Dr\k^2 - \lambda\I \right) = 0,
\]
where
$\Fr=\F(\ur)=\left(\@\f/\@\u\right)_{\u=\ur}=\Mx{f_{11}&f_{12}\\f_{21}&f_{22}}$
is the Jacobian matrix of the reaction terms and $\Dr=\D(\ur)$ is the
diffusion matrix, both evaluated at the equilibrium. Considering for
simplicity the cases of \fig{density}(a,b,d,f) where $\Dr=\Mx{0 & \hm\Pr \\ 0 &
  0}$,  we have
\[
  \lambda=\frac12\left( f_{11} + f_{22} \pm \sqrt{
      (f_{11}-f_{22})^2+4f_{12}f_{21} - f_{21} \hm\Pr\k^2
  }\right)
\]
which for
$\k^2>\max\left( ((f_{11}-f_{22})^2+4f_{12}f_{21})/(\hm\Pr) \,,\, 0\right)$
gives oscillatory behaviour of perturbations, but then
$\Re(\lambda)=(f_{11}+f_{22})/2=\const$ whereas it has to have a
maximum at a positive $\k^2$ for this mechanism to be relevant.

\paragraph{It does not emerge via Turing-Hopf instability of spatially uniform oscillations}

\sglfigure{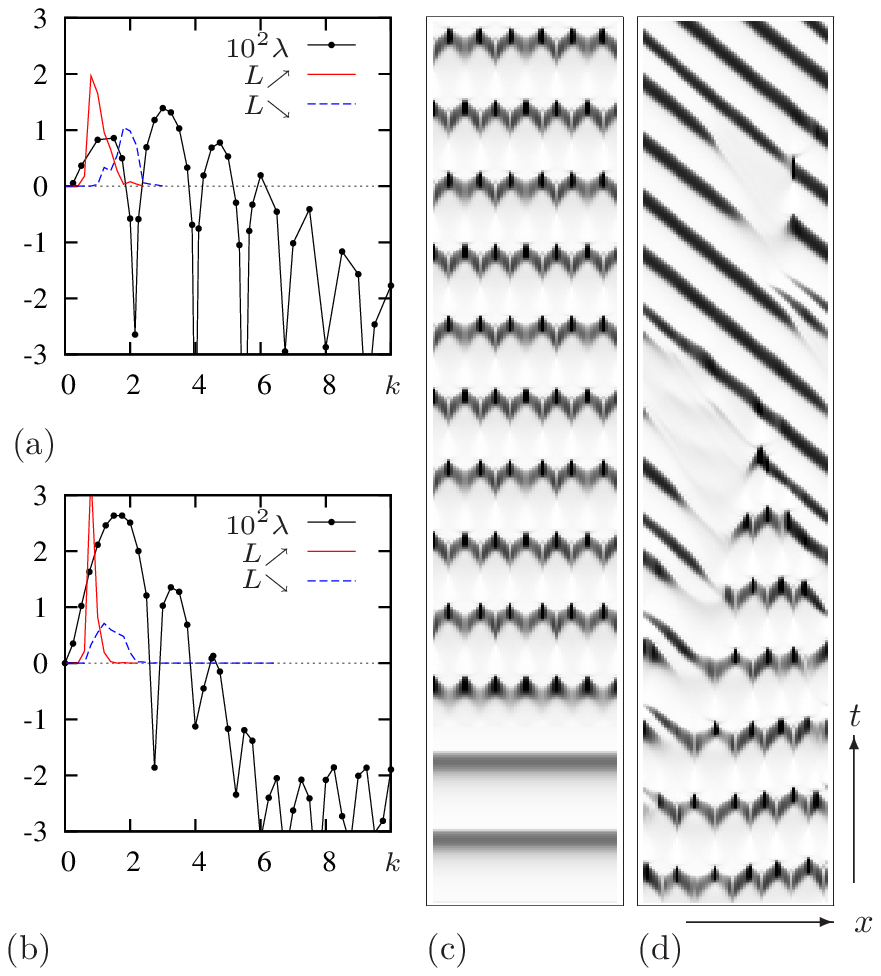}{ (color online)
  Emergence of spontaneous periodic waves through instabilities.
  (a) Spectrum $\lyap(\k)$ of harmonic perturbations of the spatially uniform
  oscillations (black lines with points), and the histograms of the
  wavenumbers of spontaneous wavetrains in the simulations shown in
  \fig{lchange} for increasing $L$ (red solid line) and decreasing $L$
  (blue dashed line).
  (b) Same, for the RM model, parameters as in \fig{density}(f),
  histograms obtained by increasing  
  $L$ by 0.2 every 2000 time units from 5 to 50 and
  decreasing it back to 2 by 0.2 every 1000 time units.  
  (c) Emergence of standing periodic waves via an instability
  of the spatially uniform oscillations. Parameters as in \fig{density}(a)
  and \fig{lchange}, $L=12.26$, $\dx=L/63$, $\dt=5\cdot10^{-5}$,
  $t\in[0,1000]$. 
  (d) Emergence of spontaneous periodic wavetrains via an instability of
  periodic standing waves. Continuation of (c), $t\in[44460,45460]$. 
}{instab}

The next possible candidate is the instability of spatially uniform
oscillations with respect to perturbations which nonzero freqency and
nonzero wavenumber. This case is not considered in the CH
nomenclature, and is called ``Hopf'' instability of spatially homogeneous
oscillations, with the same variants as in the previous case, by RS.
This instability looks plausible as spatially homogeneous (spatially uniform) oscillations in our
systems are indeed possible and even stable in small spatial
domains, so we have investigated this possibility with particular
care. As limit cycles in the point systems of~\eq{TB} and~\eq{RM} can
not be described analytically, the investigation of stability has to
be done numerically. We have considered solutions of the form
$\u(x,t)=\uo(t)+\Re\left(\v(t)e^{i\k x}\right)$ with $|\v|\ll1$, which
gives a coupled system of ordinary differential equations
\begin{subequations}
\begin{align}
  \Df{\uo}{t} &= \f(\uo),                                  \label{coupled.non} \\
  \Df{\v}{t}  &= \left( \F(\uo) - \D(\uo) k^2 \right) \v , \label{coupled.lin}
\end{align}                                                \label{coupled}
\end{subequations} 
with parameter $\k$. We solved system~\eq{coupled} forward in time
with initial conditions for bulk oscillations $\uo$ in the basin of
attraction of the limit cycle, and arbitrary nonzero initial conditions
for the perturbation $\v$. Then we estimated the Lyapunov exponent for the
$\v$-subsystem, $\lyap(\k)=\lim_{t\to\infty}t^{-1}\ln(||\v(t)||)$.
The estimation was done by finding maxima of the first
component of $\v(t)$ and linearly fitting their logarithms against
$t$, for an interval of large enough values of $t$.  For selected
values of $\k$ we used two linear independent sets of initial
conditions for $\v$, to eliminate the theoretical possibility of
accidentally chosing initial conditions that did not lead to the maximal
exponent.

The resulting graph $\lyap(\k)$ for the TB model at the same
parameters as in 
\fig{density}(a) and \figref{lchange} is shown on \fig{instab}(a). For
comparison, we also show histograms of the empirical wavenumbers
observed in simulations shown in \fig{lchange}, calculated as $\k=2\pi
n/L$, separately for the growing and decreasing $L$.
\Fig{instab}(b) shows similar graphs made for the RM model
at the same parameters as in \fig{density}(f). It is clear that,
although there are finite bands of wavenumbers producing growing
perturbations, the
actually selected wavenumbers are not the same as those of
the fastest growing perturbations, and for the TB model they even partly fall in the interval of
decaying perturbations. 

Moreover, the growing perturbations of the spatially uniform oscillations in
fact do not represent propagating periodic waves, but standing
waves. This is illustrated in \fig{instab}(c) where we show a density
plot of a simulation of the full model, similar to \fig{density}(a)
but with different initial conditions. Here we chose initial
conditions as spatially uniform oscillations plus a very small perturbation
sinusoidal in space. Note that for the limit of infinitely small
perturbation amplitudes this exactly corresponds to system
\eq{coupled}. 

We conclude that although the cross-diffusion driven instability does indeed
take place in the considered examples, the waves that emerge are in
fact quite different from the spontaneous periodic traveling waves. 

\paragraph{Spontaneous sources as a precursor of spontaneous periodic waves}

The periodic standing waves emerging via the cross-diffusion driven
instability described above, are in turn unstable themselves.
\Fig{instab}(d) shows a continuation of simulation of
\fig{instab}(c). The standing waves are observed for a long time, as
they are stable within the space of functions with spatial period
$2\pi/\k=L/6$, and the numerical initial conditions are almost exactly
periodic with that period, up to small errors resulting from finite
precision arithmetics. The small symmetry-breaking numerical errors
allow for an instability of the periodic standing waves to develop,
during which some of the standing waves occur later than others. When
this instability sufficiently develops, there is a sudden, ``hard''
transition to propagating waves. The spatial period of the propagating
waves is twice longer than the spatial period of preceding standing
waves.  We stress that the traveling waves do not appear via anything
like ``bifurcation'' from standing waves, at least in the examples we
considered.

Notice that the long transient solution shown in \fig{instab}(c,d) is
a periodic standing wave by its symmetry, but it also looks like a
periodic set of focal sources, synchronously sending out solitary
waves which then annihilate each other.  As can be seen in
\fig{density}, apart from the symmetry, this sort of transient before
the onset of periodic waves is typical, and only its duration varies
in different simulations.  That is, the special initial conditions in
\fig{instab}(c,d) only affect the symmetry and the duration of the
transient, but not its qualitative character. A similar route to
traveling waves via unstable periodic set of ``focal sources''
standing waves is obseved in the RM model.

\section{Conclusion}

The considered examples demonstrate an unusual type of behaviour. The
systems are oscillatory, but the spatially uniform oscillations are
unstable. The systems can also demonstrate standing periodic waves,
which are also unstable.  These instabilities lead to periodic
propagating waves, which seems to be the only stable regime.  This
regime emerges spontaneously even when boundary conditions disallow
propagating waves. The periods of the waves can be in a certain
interval with strict boundaries, both upper and lower. Nearer the
upper end of the interval, \ie\ at longer wavelengths, the periodic
waves do not propagate steadily but are modulated. Transition from
steady to modulated propagation is soft and has empirical features of
a supercritical Hopf bifurcation (of a relative equilibrium), \ie\
possibly an Eckhaus mechanism.

The defining features described above are sufficiently generic, and
the phenomenon of spontaneous periodic traveling waves does not
disappear as the parameters are varied, nor it is restricted just to
one model. This behaviour does not fall into existing classification
of pattern formation scenarios. The detailed mechanisms
of spontaneous generation and maintenance of periodic traveling waves
require further investigation. However, it is clear that
cross-diffusion is an essential factor, since its replacement with,
or adding of significant amount of, self diffusion
  eliminates the effect. Cross-diffusion phenomena are known in a in a
  variety of physical situations. For example, spontaneous periodic
  waves have been observed in a Burridge-Knopoff mathematical model of
  earthquakes~\cite{Cartwright-etal-1997,Cartwright-etal-1999}. That
  model belongs to the class \eq{RD}, with only one nonzero element
  of matrix $\D$, as in our simulations shown in~\fig{density}(a) and
  (g) but constant, and
  excitable FitzHugh-Nagumo local kinetics. It is not known whether
  the spontaneous waves in the Burridge-Knopoff model have a finite
  interval of allowed wavenumbers, as illustrated by~\fig{lchange} for
  our case, however other described features of those waves are
  similar to those described here and are likely to have a
  similar nature. Further investigation of the mechanism of generation
  of such waves is a subject for further study which is of
broad physical interest as a new pattern forming mechanism in
dissipative spatially distributed systems.

Returning to the application that originally motivated this study,
attempts to explain waves observed in cyclic biological populations,
using reaction-diffusion models, had to involve spatially-nonuniform
external factors, \eg\ sites of increased mortality due to
environmental conditions
\cite{Sherratt-etal-2002,Sherratt-Smith-2008}. Such factors are needed
to disallow uniform oscillations. Our present results imply that such
factors may not be necessary if cross-diffusion interaction are taken
into account as the uniform oscillations may be unstable and waves
form spontaneously.

\textbf{Acknowledgments}
We are grateful to O.~Piro for helpful advice.
The study was supported in part by RFBR grant 07-04-00363 (Russia) and by a
grant from the Research Centre for Mathematics and Modelling of
Liverpool University (UK).


\end{document}